\documentclass{iau}

\usepackage{amsmath}
\usepackage{graphicx}
\usepackage{multirow}

\newcommand{\oiii}{[O{\sc iii}] }

\begin{document}

\lefttitle{N. Chornay}
\righttitle{Probing the Local Planetary Nebula Luminosity Function with \textup{Gaia}}

\journaltitle{Planetary Nebulae: a Universal Toolbox in the Era of Precision Astrophysics}
%\jnlPage{1}{7}
\jnlDoiYr{2023}
\doival{10.1017/xxxxx}
\volno{384}

\aopheadtitle{Proceedings IAU Symposium}
\editors{O. De Marco, A. Zijlstra, R. Szczerba, eds.}
 
\title{Probing the Local Planetary Nebula\\Luminosity Function with \textit{Gaia}
}

\author{Nicholas Chornay$^{1,2}$, Nicholas Walton$^2$, David Jones$^{3,4}$, and Henri Boffin$^5$}
\affiliation{$^1$ Department of Astronomy, University of Geneva, Chemin d'Ecogia 16, 1290 Versoix, Switzerland;\\
email: {\tt Nick.Chornay@unige.ch}\\
$^2$ Institute of Astronomy, University of Cambridge, Madingley Road, Cambridge CB3 0HA, UK \\
$^3$ Instituto de Astrof\'isica de Canarias, E-38205 La Laguna, Tenerife, Spain \\
$^4$ Departamento de Astrof\'isica, Universidad de La Laguna, E-38206 La Laguna, Tenerife, Spain \\
$^5$ ESO, Karl-Schwarzschild-str. 2, 85748 Garching, Germany
}

\begin{abstract}
The Planetary Nebula Luminosity Function (PNLF) remains an important extragalactic distance indicator despite a still limited understanding of its most important feature - the bright cut-off. External galaxies benefit from consistent distance and extinction, which makes determining the PNLF easier but detailed study of individual objects much more difficult. Now, the advent of parallaxes from the Gaia mission has dramatically improved distance estimates to planetary nebulae (PNe) in the Milky Way. We have acquired ground-based narrowband imagery and measured the [OIII] fluxes for a volume-limited sample of hundreds of PNe whose best distance estimates from Gaia parallaxes and statistical methods place them within 3~kpc of the Sun. We present the first results of our study, comparing the local PNLF to other galaxies with different formation histories, and discussing how the brightness of the PNe relates to the evolutionary state of their central stars and the properties of the nebula.

\end{abstract}


\maketitle

\section{Introduction}

The \oiii $\lambda5007$ magnitudes of planetary nebulae (PNe) show a remarkably consistent bright-end cutoff largely independent of galactic age and metallicity \citep{ciardullo1989, jacoby1989}, so much so that the \oiii planetary nebula luminosity function (PNLF) continues to serve as an important extragalactic distance indicator \citep{roth2021, scheuermann2022}. The reason for this behaviour is, however, still not fully understood after more than three decades \citep{ciardullo2022}.

Historically, highly uncertain distances to Galactic PNe \citep{frewsurfacebrightness2016} have made the study of the PNLF of the Milky Way difficult, outside of populations at a known (assumed) distance such as the Galactic Bulge \citep{kovacevic2011fluxes}. Milky Way PNLF studies are also challenged by the presence of interstellar dust, whose effects must be disentangled from those of circumnebular dust \citep{davis2018pnlfextinction,yao2023pnlforigin} in order to recreate the PNLF's behaviour. These issues mean that studies of the PNLF have largely focused on other galaxies \citep[e.g. M31:][]{bhattacharya2019,galera2022}, where all PNe can be treated as lying at the same distance (not to mention more efficiently imaged). However a unique opportunity of studying the Milky Way PNLF is the direct observability of their CSPNe, which can be probed in detail for indicators of stellar parameters and possible binary evolution. The morphologies of the nebulae are also visible, spectroscopy is easier to obtain, and there is already a wealth of information in the literature.

The advent of precise parallaxes from the \textit{Gaia} mission \citep{gaiamission2016} has dramatically improved distance estimates to Galactic PNe \citep{chornaywalton2021}, and \textit{Gaia} data has also been used to improve Milky Way extinction maps \citep[see, e.g.,][]{dharmawardena2022oriondust}. A missing ingredient is \oiii $\lambda5007$ flux measurements. While it is possible to estimate \oiii $\lambda5007$ fluxes using fluxes in other lines and spectroscopic line ratios, that approach introduces significant systematic uncertainties, particularly for nearby objects with a greater angular extent. The best values come from direct measurement \citep[see e.g. the assessment of ][]{kovacevic2011fluxes}.

In this work we introduce our narrowband imaging survey of local PNe within a 3~kpc heliocentric radius. We use those fluxes, combined with extinction and distance estimates, to determine the PNLF of the sample. We compare this PNLF to those of other nearby galaxies, and describe how the properties of the nebulae and the CSPNe relate to the absolute \oiii magnitudes of the PNe.

\section{Observations and Data Reduction}

Targets were selected with the aim of obtaining fluxes for a 3~kpc volume-limited sample of PNe. Targets were "true" (spectroscopically confirmed) PNe from the Hong Kong/AAO/Strasbourg H$\alpha$ (HASH) PN catalogue \citep{hashpn}. The distance estimates used for selection were the median distances from the catalogue of \citet{chornaywalton2021} in the cases where there was a sufficiently secure central star identification. PNe were also included if their statistical distance estimate from \citet{frewsurfacebrightness2016} or the inversion of the (sufficiently high quality) \textit{Gaia} parallax placed them within 3~kpc.

Data for targets visible from the northern hemisphere (northwards of $-25^{\circ}$ in declination) were obtained with the Wide Field Camera (WFC) on the 2.54m Isaac Newton Telescope (INT) and the Alhambra Faint Object Spectrograph and Camera (ALFOSC) on the 2.56m Nordic Optical Telescope (NOT), both at the Observatorio del Roque de los Muchachos in La Palma, Spain. Southern hemisphere targets were observed at the European Southern Observatory (ESO) New Technology Telescope (NTT) at the La Silla observatory in Chile, using the ESO Faint Object Spectrograph and Camera (EFOSC2). That latter set of observations are not presented here.

Across all instrument setups, targets were observed with a narrowband filter centred around the \oiii $\lambda$5007 emission line as well as a broad filter whose transmission curve enclosed that of the narrow filter (nominally a $g$ filter). Typical exposure times were 300 seconds in the narrowband and 30 seconds in the broadband filter, both repeated twice; these were shortened for very bright targets to avoid saturation.

Images were reduced using standard methods. Sources were extracted from each reduced image using the {\sc imcore} program published by the Cambridge Astronomical Software Unit (CASU).\footnote{\url{http://casu.ast.cam.ac.uk/surveys-projects/software-release}}, and a world coordinate system (WCS) was determined by matching those sources to the \textit{Gaia} EDR3 catalogue, propagated to the epoch of the observations.

PN fluxes were extracted in each \oiii image using a circular aperture, whose size was adjusted to match that of the PN. The background was estimated using an annulus centred on the PN. Particularly for fainter nebulae, field stars coincident with the nebula can contribute significantly to the aperture flux; this was removed by masking the stars, using the \textit{Gaia} positions and magnitudes as a reference.

Photometric calibration was bootstrapped from the stellar fluxes of the near contemporaneous g-band image, which were compared to fluxes from the Pan-STARRS survey (for the northern hemisphere data). The nominal central wavelengths of the various g-band filters are all close to that of the \oiii filters, so we expect differences in the atmospheric extinction between these bands to be relatively small and predictable. The relationships between the zero points in the two filters as a function of airmass were determined from standard star observations included in each night of observing.

The uncertainties are dominated by effects other than photon counting: systematic errors due to the image reduction, flux extraction, and the method of determining the zero point, as well as further uncertainties due to potential variations of the zero point of the course of observations. Empirical uncertainties were obtained from repeat exposures. The conversion of narrowband filter to line flux depends on the transmission of the narrowband filters and radial velocity shifts. The former is only be detectable through systematic flux differences between instruments or between our measurements and those from literature; the latter we expect to be small and ignorable for our purposes, with the range of radial velocities of objects in our sample observed by \citet{durand1998kinematics} resulting in a transmission change of at most about 1\% in the narrowest \oiii filter, that of the NOT.

\section{\oiii Fluxes}

We observed 225 different PNe, of which 219 had positive flux measurements, and 216 had positive flux measurements with a signal-to-noise ratio greater than one. The resulting fluxes span five orders of magnitude.

\begin{figure}
\begin{center}
    \includegraphics[width=0.48\columnwidth]{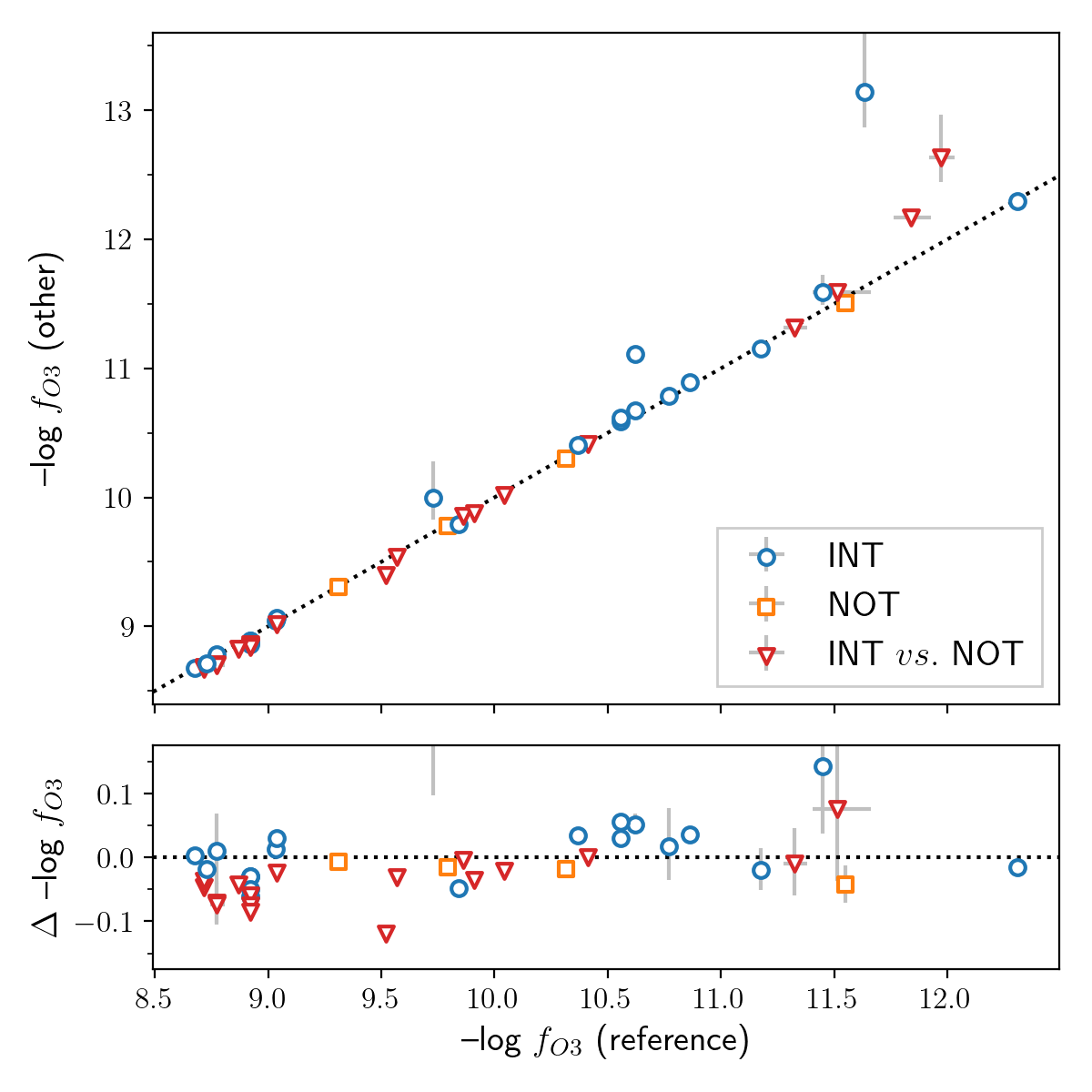}
    \includegraphics[width=0.48\columnwidth]{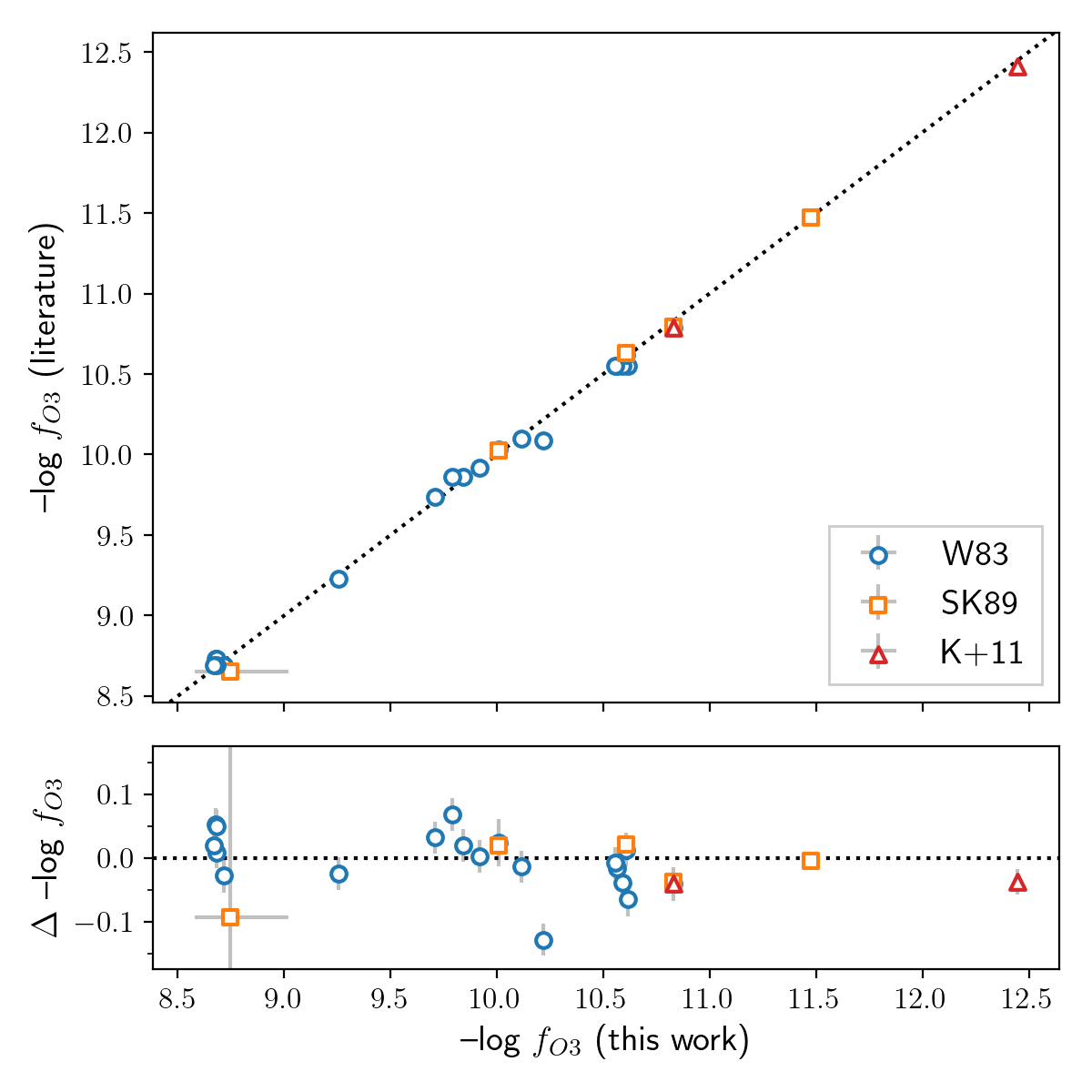}
\end{center}
\caption{Left: Comparison of log fluxes for targets with repeated measurements, for targets observed by the INT, NOT, and both. The reference value is the observation with the best nominal uncertainties (grey error bars), except for the comparison between the INT and NOT fluxes in which case it is the INT observation with the best nominal uncertainties. Right: Comparisons to previous directly measured fluxes in literature, from \citet{webster1983elfluxes} (W83), \citet{shawkaler1989oiifluxes} (SK89), and \citet{kovacevic2011fluxes} (K+11).}
\label{fig:flux_repeats}
\end{figure}

We repeated observations of some of the same targets on different nights and some with different instruments. The purpose of this is both to empirically assess the uncertainty and to look for systematic differences in our calibration between instruments. The results are shown in the left panel of Fig.~\ref{fig:flux_repeats}. The INT measurements have a scatter of 0.05 dex, while the four NOT measurements show about half that. For targets observed with both instruments, the NOT observations tend to be slightly brighter than the INT ones.

A small number of our targets have direct \oiii flux measurements in literature that can be used for comparison. This was by design, as few nearby northern PNe have such measurements. We included two targets overlapping with \citet{kovacevic2011fluxes}, as well as five from \citet{shawkaler1989oiifluxes} and thirteen from \citet{webster1983elfluxes}. The results of these comparisons are shown in the right panel of Fig.~\ref{fig:flux_repeats}; all of these were measured at the INT, except for the two points closest the the equality line at the bright end which came from the NOT. There is not a significant systematic offset between our flux measurements and those in literature, except that both of the fluxes of \citet{kovacevic2011fluxes} are brighter than ours by 0.04 dex.

While the measurements certainly show repeatability, in general the scatter is larger than expected from the nominal uncertainties, though the obvious outliers do have noticeably larger error bars. More work is needed to disentangle the source of this additional error: whether it is due to the measurement of the nebula flux itself, or uncertainty in the zero point, or some other systematic error introduced in the data reduction process. The comparison to literature fluxes show neither an obvious offset nor a higher scatter than expected from the internal consistency checks, so we consider the internal consistency to be the first priority for improvement.

\section{PNLF}

\begin{figure}
\begin{center}
    \includegraphics[width=0.8\columnwidth]{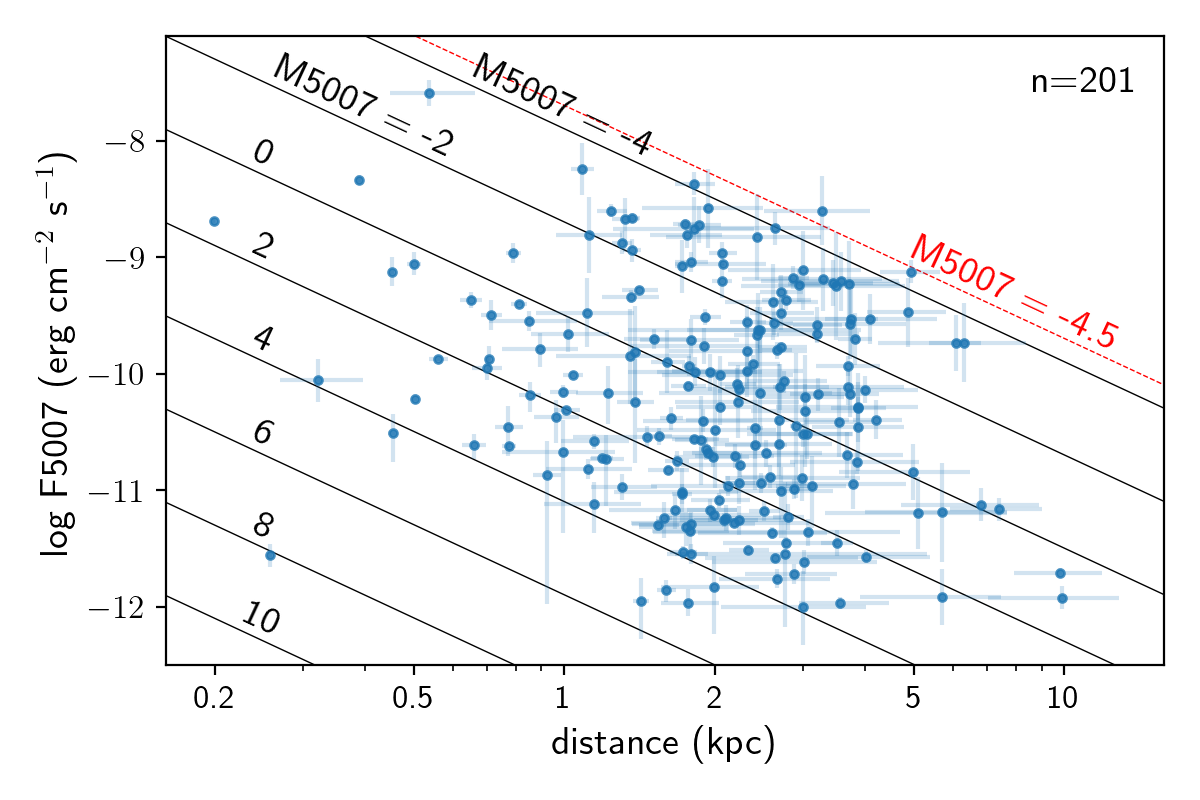}
\end{center}
\caption{Logarithmic \oiii fluxes corrected for extinction, for all observed PNe with flux signal-to-noise greater than one and literature extinction estimates. The diagonal lines show different values of $M_{5007}$ with the approximate cutoff magnitude highlighted as the dashed red line.}
\label{fig:fluxdist}
\end{figure}

The distribution of de-reddened fluxes with distance is shown in Fig.~\ref{fig:fluxdist}, using the median distance estimates from \citet{chornaywalton2021} and extinction values from \citet{frewsurfacebrightness2016}. For the purposes of this initial analysis we do not attempt to separate interstellar extinction and self-extinction by the PNe. The sample is nearly complete to 3~kpc (compared to the sample of known, spectroscopically confirmed PNe in the survey area accessible from the Northern hemisphere); we use this as the cutoff distance for most of the subsequent analysis.

The diagonal lines show corresponding values of $M_{5007}$; its distribution spans 8 magnitudes but is naturally incomplete at the faint end. The uncertainties in $M_{5007}$ are taken from the uncertainties corresponding to flux, distance, and extinction added in quadrature.

\begin{figure}
\begin{center}
    \includegraphics[width=0.8\columnwidth]{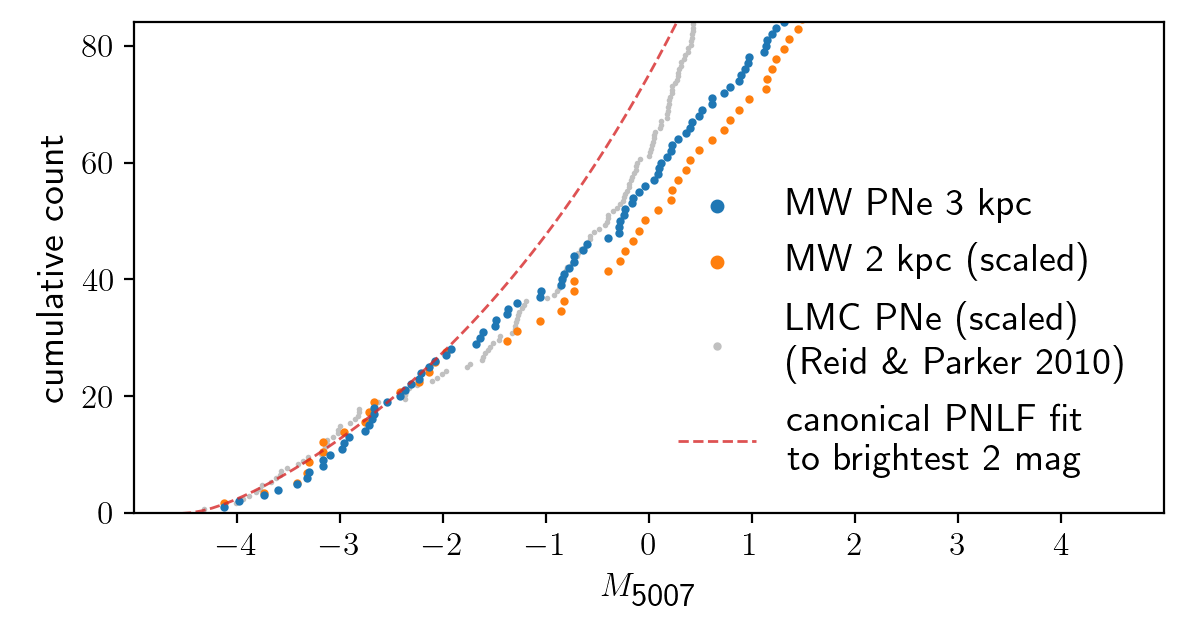}
\end{center}
\caption{Cumulative distribution of $M_{5007}$ values for our 3~kpc Milky Way sample compared to those of the same sample limited to 2~kpc and of the LMC sample of \citet{reidparker2010lmcpnlf}. The latter samples have been scaled to match the count in the brightest magnitude bins of the 3~kpc sample. The red dashed line is the canonical PNLF form scaled to match the counts at the bright end.}
\label{fig:pnlf}
\end{figure}

The resulting cumulative distribution of $M_{5007}$ with the 3~kpc sample is shown in Fig.~\ref{fig:pnlf}. It is compared to the canonical distribution from \citet{ciardullo1989}, $ N(M) \propto e^{0.307 M} (1 - e^{3 (M^* - M)}) $, 
with the cutoff magnitude fixed at $M^*=-4.5$ and the total count scaled to match the total number of PNe in the sample with $M_{5007} < -2.5$. The choice of scale is motivated by the clear departure of the distribution from the canonical one at fainter magnitudes. This corresponds to a dip in the (non-cumulative) PNLF.

Such a feature can be present in other galaxies that have undergone more recent star formation \citep{ciardullo2022}. For example, the Small Magellanic Cloud (SMC) PNLF of \citet{jacobydemarco2002smcpnlf} also shows a dip starting about 2.5 magnitudes below the bright cutoff, while the dip in the Large Magellanic Cloud (LMC) PNLF of \citet{reidparker2010lmcpnlf} occurs even closer to the cutoff magnitude (grey points in Fig.~\ref{fig:pnlf}). \citet{jacobydemarco2002smcpnlf} speculate that this gap is associated with a phase of fast CSPN evolution, in particular in young populations, noting that, if that is indeed the case, then the position of the dip is related to timing of the most recent period of star formation. We caution that the PNLF presented here is not directly comparable to these PNLFs, because the fluxes are corrected for extinction.

The PN in our sample with the brightest measured \oiii flux is NGC 6572, with $M_\textup{5007} = -4.1\pm0.3$. It has a bipolar morphology and is classified as a Type IIa PN by \citet{quireza2007peimbert}.

The morphologies of the PNe (taken from HASH PN) show a small preference for bipolarity at the bright end, but otherwise bipolar and elliptical PNe are distributed relatively evenly (notably it is more difficult to measure the morphologies of fainter PNe, so the classifications are biased). This is in contrast with the observation of \citet{kovacevic2011bulge}, who noted that bipolar PNe preferentially occupied the faint end of their Bulge PNLF (though without much detail). Round PNe are only present at magnitudes fainter than the dip.

Most known close binary systems occupy the intermediate magnitudes of the PNLF. The brightest known binary is NGC 2392 ($M_\textup{5007}=-2.4\pm0.3$), a double-degenerate system detected through radial velocity variations \citep{miszalski2019ngc2392}. NGC 6572 has bipolar outflows \citep{miranda1999outflowsngc6572}, which are often associated with binary interactions \citep{boffinjones2019}, but the brightness of the object (and indeed of the brightest PNe) makes it difficult to monitor for photometric or radial velocity variations from the ground, so its status has not been conclusively shown.

\section{Conclusions}
When published our catalogue will represent the largest set of directly measured \oiii fluxes for nearby (non-Bulge) Milky Way PNe, spanning nearly five orders of magnitude in flux and with typical uncertainties of 0.04 dex. The data already shown here will be complemented by recently acquired observations from the Southern hemisphere, and will provide a rich basis for future analysis and studies of Milky Way PNe, more so when combined with more precise parallax measurements from future \textit{Gaia} data releases.
\\

This research is based on observations made with the Isaac Newton Telescope operated on the island of La Palma by the Isaac Newton Group of Telescopes in the Spanish Observatorio del Roque de los Muchachos of the Instituto de Astrofísica de Canarias. This research is based on observations made with the Nordic Optical Telescope, owned in collaboration by the University of Turku and Aarhus University, and operated jointly by Aarhus University, the University of Turku and the University of Oslo, representing Denmark, Finland and Norway, the University of Iceland and Stockholm University at the Observatorio del Roque de los Muchachos, La Palma, Spain, of the Instituto de Astrofisica de Canarias. This research has made use of data from the European Space Agency (ESA) mission \textit{Gaia} (\url{https://www.cosmos.esa.int/gaia}), processed by the {\it Gaia} Data Processing and Analysis Consortium (DPAC; \url{https://www.cosmos.esa.int/web/gaia/dpac/consortium}). Funding for the DPAC has been provided by national institutions, in particular the institutions participating in the {\it Gaia} Multilateral Agreement. This research has also made use of the HASH PN database (\url{http://hashpn.space}). This research was supported through the Cancer Research UK grant A24042.

% \bibliographystyle{iaulike.bst}
% \bibliography{bib}

\end{document}